\documentclass[twocolumn,english,conference]{IEEEtran}
\usepackage[T1]{fontenc}
\usepackage{geometry}
\usepackage{babel}
\usepackage{float}
\usepackage{amsthm}
\usepackage{amsmath}
\usepackage{amssymb}
\usepackage{graphicx}
\usepackage[unicode=true,
 bookmarks=true,bookmarksnumbered=true,bookmarksopen=true,bookmarksopenlevel=1,
 breaklinks=false,pdfborder={0 0 0},backref=false,colorlinks=false]
 {hyperref}
\hypersetup{pdftitle={Your Title},
 pdfauthor={Your Name},
 pdfpagelayout=OneColumn, pdfnewwindow=true, pdfstartview=XYZ, plainpages=false}

\makeatletter

\floatstyle{ruled}
\newfloat{algorithm}{tbp}{loa}
\providecommand{\algorithmname}{Algorithm}
\floatname{algorithm}{\protect\algorithmname}

\theoremstyle{plain}
\newtheorem{thm}{\protect\theoremname}
\theoremstyle{definition}
\newtheorem{defn}[thm]{\protect\definitionname}
\theoremstyle{plain}
\newtheorem{lem}[thm]{\protect\lemmaname}
\theoremstyle{plain}
\newtheorem{cor}[thm]{\protect\corollaryname}
\theoremstyle{plain}
\newtheorem{prop}[thm]{\protect\propositionname}

\usepackage[caption=false,font=footnotesize]{subfig}

\makeatother

\providecommand{\corollaryname}{Corollary}
\providecommand{\definitionname}{Definition}
\providecommand{\lemmaname}{Lemma}
\providecommand{\propositionname}{Proposition}
\providecommand{\theoremname}{Theorem}

\begin{document}

\title{Asymptotically Optimal Distributed Channel Allocation: a Competitive
Game-Theoretic Approach}

\author{\IEEEauthorblockN{Ilai~Bistritz}\IEEEauthorblockA{Department of Electrical Engineering-Systems\\
Tel-Aviv University, Israel\\
Email: ilaibist@gmail.com}\and \IEEEauthorblockN{Amir~Leshem}\IEEEauthorblockA{Faculty of Engineering\\
Bar-Ilan university, Ramat-Gan, Israel\\
Email: leshema@eng.biu.ac.il}}
\maketitle
\begin{abstract}
In this paper we consider the problem of distributed channel allocation
in large networks under the frequency-selective interference channel.
Performance is measured by the weighted sum of achievable rates. First
we present a natural non-cooperative game theoretic formulation for
this problem. It is shown that, when interference is sufficiently
strong, this game has a pure price of anarchy approaching infinity
with high probability, and there is an asymptotically increasing number
of equilibria with the worst performance. Then we propose a novel
non-cooperative M Frequency-Selective Interference Game (M-FSIG),
where users limit their utility such that it is greater than zero
only for their M best channels, and equal for them. We show that the
M-FSIG exhibits, with high probability, an increasing number of optimal
pure Nash equilibria and no bad equilibria. Consequently, the pure
price of anarchy converges to one in probability in any interference
regime. In order to exploit these results algorithmically we propose
a modified Fictitious Play algorithm that can be implemented distributedly.
We carry out simulations that show its fast convergence to the proven
pure Nash equilibria. 
\end{abstract}

\section{Introduction}

Channel allocation, the problem of assigning frequency bands to users,
is a fundamental element in wireless networks. Channel allocation
is necessary when channel access is through frequency division techniques
such as FDMA or the more recent bandwidth efficient technique OFDMA.
Other approaches, based on iterative water filling (IWF, see \cite{Yu2002}),
allow users to allocate their power over the spectrum as a whole.
It is well-known that IWF leads to a FDMA solution for strong interference,
and hence is more suitable for weak interference and is generally
considered more complex. When splitting the channel into sub-channels,
the question of how to assign these sub-channels to users arises.
In the frequency-selective interference channel, different users experience
different conditions in each channel due to fading and interference,
so different allocations will result in varying levels of performance.

At first glance, it may seem that channel allocation is a special
case of resource allocation and as such can be solved as an optimization
problem. If $N$ is the number of users and resources, the optimal
permutation between them can be found with a complexity of $O(N^{3})$,
using the famous Hungarian Algorithm \cite{Papadimitriou98}. The
basic problem with this approach is the information required to compute
the optimal solution. To do so, some network entity (the base station,
access point, etc) needs to know the preferences for all nodes. This
entity should compute the optimal solution and transmit it back to
the nodes. In a wireless environment, these preferences are not constant
so this central knowledge involves significant communication overhead
on the network. As networks grow larger, this requirement becomes
less reasonable. Furthermore, future networks (like ad-hoc networks
and cognitive radio) are envisioned to be more distributed in nature
and less dependent on central entities. This leads to the need for
a distributed channel allocation algorithm.

Recently, it has been shown that the optimal solution to the channel
allocation problem can be achieved using a distributed algorithm \cite{Naparstek2013,Naparstek2014,Naparstek2014a}.
This algorithm is a distributed version of the auction algorithm \cite{Bertsekas79}
and relies on a CSMA protocol. Although it has a very slow convergence
rate, this result serves as a proof of concept and suggests that other
approaches may achieve close to optimal performance in a distributed
fashion. In \cite{Leshem2012}, the authors designed an algorithm
based on the stable matching concept that also uses a CSMA protocol.
This algorithm has a much faster convergence rate and a good sum-rate
performance. Due to their dependence on CSMA, both algorithms are
vulnerable to the hidden terminal and exposed terminal problems. In
order to avoid these problems, a RTS/CTS mechanism has to be implemented.
Besides causing delays, RTS/CTS implementation requires some central
network entities, and thus negatively impact the network scalability.
Additionally, both algorithms have strong user synchronization requirements.
Last, but not least, these algorithms ignore the inherent possibility
of sharing channels between users.

There has been a considerable amount of work designed to apply game
theory as a framework for distributed channel allocation algorithms
(see \cite{Larsson2009,Scutari2008}). While game theory addresses
the distribution requirement naturally, it does not guarantee good
global performance. For example, it is well-known that the fixed points
for the IWF algorithm are the Nash equilibrium points of the Gaussian
interference game. For a two-user Gaussian interference game, a prisoner’s
dilemma may occur which leads to a suboptimal solution \cite{Laufer2005}.
To overcome this obstacle, some form of cooperation can be added using
different game theoretic concepts. In \cite{Nie2005} the authors
proposed a potential game theoretic formulation that requires each
user to know the interference he causes to other users. In \cite{Han2005}
the authors used the Nash bargaining solution and coalitions to enhance
the fairness of the allocation at the price of a centralized architecture.
In \cite{Leshem2008} and \cite{Leshem2011}, a more stable algorithm
to obtain the Nash bargaining solution was proposed, based on convex
optimization techniques. Although cooperation can indeed enhance performance,
it may be extremely complicated to achieve cooperative game-theoretic
solution concepts in the general case without communication between
users, which limits the distributed nature of the network. 

The rest of this paper is organized as follows. In section II we formulate
our wireless network scenario and present our approach. In section
III we present a natural game formulation for this problem and show
that it suffers from major drawbacks. In section IV we propose an
enhanced game and provide its equilibria analysis. Section V suggests
an algorithm each user can implement in a distributed fashion to converge
to these equilibria. In section VI we present simulations of our proposed
algorithm that show fast convergence to the proven equilibria. Finally,
we draw conclusions in section VII.

\section{Problem Formulation}

Consider a wireless network consisting of $N$ transmitter-receiver
pairs (users) and $K$ frequency bands (channels). Each user forms
a link between his transmitter and receiver using a single frequency
band. The channel between each transmitter and receiver is Gaussian
frequency-selective and we assume that each frequency band is smaller
than the coherence bandwidth of the channel. We also assume that the
coherence time is large enough so that the channel gains can be considered
static for a sufficiently long time.

The channel gains (fading coefficients) are modeled as $N{}^{2}K$
independent random variables - one for each channel, each transmitter
and each receiver. The coefficient between user $n$s transmitter
and user $m$s receiver in channel $k$ is denoted $h_{n,m,k}$. We
also assume that $h_{m,n,1},...,h_{m,n,K}$ are identically distributed
for each $m,n=1,...,N$.

Note that $NK$ of these coefficients serve as channel coefficients
between a transmitter and receiver pair and are denoted for convenience
by $h_{n,k}$ for user $n$ in channel $k$. The other $NK(N-1)$
coefficients serve as interference coefficients between transmitters
and unintended receivers. In this paper we assume $N=K$ for simplicity.

Each user has some preferred order of the $K$ channels. Due to the
independence of the channel coefficients between users, these preference
lists are different and independent between users. Note that this
preference order considers only the absolute value of the channel
coefficient and not the interference (which indeed affects the achievable
rate). We denote by $h{}_{n,(N-i+1)}$ the i-th best channel coefficient
for user $n$ (so $h{}_{n,(1)}$ is the worst channel). 

We assume that each user has perfect channel state information (CSI)
of all his $K$ channel coefficients, which he can achieve using standard
estimation techniques. In addition we assume that each user can sense
the exact interference he experiences in each channel. Nevertheless,
users do not have any knowledge about the channel coefficients of
other users or about any of the interference coefficients. There is
no central entity of any sort that knows the channel coefficients
of all users. Note that, in contrast to \cite{Naparstek2014} and
\cite{Leshem2012}, we do not prohibit two or more users in the same
channel.

Our global performance metric is the weighted sum of achievable rates
while treating interference as noise. Denote by $\mathbf{a}$ the
allocation vector (soon to be called the strategy profile), s.t. $a_{n}=k$
if user $n$ is using channel $k$. We want to maximize the following
performance function over all possible allocations
\[
W(\mathbf{a})=\sum_{n=1}^{N}w_{n}\log_{2}\left(1+\frac{P_{n}|h_{n,a_{n}}|^{2}}{N_{0}+I_{n,a_{n}}\left(\mathbf{a}_{-n}\right)}\right)
\]
where $N_{0}$ is the Gaussian noise variance which is assumed to
be the same for all users, $P_{n}$ is user $n$s transmission power
and $I_{n,k}\left(\mathbf{a}_{-n}\right)=\underset{m|a_{m}=k}{\sum}|h_{m,n,k}|^{2}P_{m}$
is the interference user $n$ experiences in channel $k$. We assume
that the weights satisfy $w_{min}\leq w_{n}\leq w_{max}$ for some
$w_{min},w_{max}>0$, for all $n$. 

We want to find a fully distributed way to achieve close to optimal
solutions for our channel allocation problem. Hence we need to analyze
the interaction that results from $N$ independent decision makers
and ensure that the outcome is desirable. The natural way to tackle
this problem is by applying game theory. 
\begin{defn}
A normal-form game is defined as the tuple 
\[
G=<\mathcal{N},\{A_{n}\}_{n\in\mathcal{N}},\{u_{n}\}_{n\in\mathcal{N}}>
\]
 where $\mathcal{N}$ is the set of players, $A_{n}$ is the set of
pure strategies of player $n$ and $u_{n}:A_{1}\times...\times A_{N}\rightarrow\mathbb{R}$
is the utility function of player $n$.
\end{defn}
Game theory aims at analyzing the possible outcomes of a given interaction
using solution concepts. The best known solution concept is the celebrated
Nash Equilibrium (NE).
\begin{defn}
A strategy profile $(a_{n}^{*},\mathbf{a}_{-n}^{*})\in A_{1}\times...\times A_{N}$
is called a pure Nash equilibrium (PNE) if $u_{n}(a_{n}^{*},\mathbf{a}_{-n}^{*})\geq u_{n}(a_{n}\mathbf{,\mathbf{a}_{-n}^{*}})$
for all $a_{n}\in A_{n}$ and all $n\in\mathcal{N}$.
\end{defn}
This means that for each player $n$, if the other players act according
to the equilibrium, player $n$ can not improve his utility with another
strategy. A game may exhibit multiple pure NE or none at all. 

A more general notion of an equilibrium is the mixed Nash equilibrium,
which is a probability assignment on the pure strategies set. It is
well known that in any game with a finite number of players and finite
strategy spaces, there exists a mixed NE \cite{Nash1951}. We choose
to avoid the notion of mixed NE due to its lack of practical meaning
as a solution for the channel allocation problem.

In our case, the players are users (through their receiver) and the
set of strategies for each player is the set of channels. The choice
of the utility function is a more delicate issue. One of our goals
in this work is to show that this degree of freedom in the choice
of the utility function can be exploited to achieve better global
performance without inducing coordination between the users. Thus
we distinguish between the global performance metric and the utility
function each user aims to maximize, and we view the dynamic of the
game solely as an algorithmic tool to converge to the desired steady
state point (NE) in terms of global performance.

Unfortunately, not every game formulation has nice equilibria in terms
of both tractability and performance. The notion of NE helps us predict
the outcome of the resulting interaction between programmed distributed
agents. The problem of tuning the dynamics to a desired equilibrium
among all existing NE (equilibrium selection) is generally difficult
and may require some coordination between the users. For this reason,
a game formulation that results in a simple and robust equilibrium
is desirable. The cost of this uncertainty on the resulting NE is
often measured by the price of anarchy, defined as follows.
\begin{defn}
The pure price of anarchy (PPoA) of a game $G=<\mathcal{N},\{A_{n}\}_{n\in\mathcal{N}},\{u_{n}\}_{n\in\mathcal{N}}>$
with the performance function $W:A_{1}\times...\times A_{N}\rightarrow\mathbb{R}$
is $\frac{\underset{\mathbf{a}\in A_{1}\times...\times A_{N}}{\max}W(\mathbf{a})}{\underset{\mathbf{a}\in E_{p}}{\min}W(\mathbf{a})}$,
where $E_{p}$ is the set of PNE.
\end{defn}
It is not hard to think of special cases of interference networks
that have bad equilibria or no pure equilibria at all. We are interested
in the vast majority of networks as dictated by the fading distribution,
especially in large networks. Therefore, our approach is probabilistic
and asymptotic in the number of users $N$ (i.e. will produce results
in the ``with high probability'' sense).

\section{The Naive Frequency-Selective Interference Game }

Given our performance metric, a natural choice for the utility of
each user is his achievable rate. This choice makes the weighted sum-rate
the weighted social welfare of the game. This means that in this game
we do not exploit the degree of freedom when choosing the utility
function and hence we call this the ``naive game''. This naiveté
can be interpreted as selfishness of the users and we will show that
it may lead to poor global performance.
\begin{defn}
\textbf{The Naive Frequency-Selective Interference Game (Naive-FSIG)}
is a normal-form game with $N$ users as players, where each has $A_{n}=\left\{ 1,2,...,K\right\} $
as a strategy space. The utility function for player $n$ is 
\[
u_{n}\left(\mathbf{a}\right)=\log_{2}\left(1+\frac{P_{n}|h_{n,a_{n}}|^{2}}{N_{0}+I_{n,a_{n}}\left(\mathbf{a}_{-n}\right)}\right)
\]

\end{defn}
In this section we analyze the PNE of the Naive-FSIG for strong interference
and evaluate the PPoA. Trivially, a user who obtained his best channel
without interference cannot improve his utility. On the other hand,
a user who is not in his best channel (with the best channel coefficient)
cannot necessarily improve his utility if there are users in his more
preferable channels. The influence of other users in the channel of
user $n$ on his utility is caused by the interference. Consequently,
the strength of the interference has a crucial effect on the identity
of the NE. 

If the interference is strong enough, users in the same channel achieve
negligible utility and the interference game becomes a ``collision
game''. 
\begin{lem}
\label{NE are Permutations}If $\frac{1}{N_{0}}\underset{k}{\min}\left|h_{n,k}\right|^{2}>\underset{l}{\max}\frac{\left|h_{n,l}\right|^{2}}{N_{0}+\underset{m}{\min}\left(\left|h_{m,n,l}\right|^{2}P_{m}\right)}$
for each $n$, then the set of PNE of the Naive-FSIG is the set of
permutations between users and channels, with cardinality $N!$. \end{lem}
\begin{IEEEproof}
The inequality condition means that for every strategy profile that
is a permutation of users to channels, a user who deviates gets lower
utility. Consequently, every permutation is an equilibrium. Conversely,
every pure equilibrium must be a permutation because all users prefer
an empty channel over a shared one (i.e. with positive interference). 
\end{IEEEproof}
The lemma above implies that in strong enough interference, a PNE
of the Naive-FSIG may assign some users a bad channel. The next lemma
shows that a bad channel can be asymptotically useless. 
\begin{lem}
\label{Worst Channels are Asymptotically Worthless}Assume that $\left|h_{n,1}\right|,\,...,\,\left|h_{n,N}\right|$
are i.i.d for each $n$, with continuous distribution $F_{n}\left(x\right)$,
s.t. $F_{n}\left(x\right)>0$ for all $x>0$. Let $M_{N}$ be a sequence
s.t. $\underset{N\rightarrow\infty}{\lim}\frac{M_{N}}{N}=0$. If $m\leq M_{N}$
then $\underset{n}{\max\,}\left|h_{n,(m)}\right|\rightarrow0$ in
probability as $N\rightarrow\infty$.\end{lem}
\begin{IEEEproof}
Let $\varepsilon>0$. Due to the i.i.d assumption, the number $N_{\varepsilon,n}$
of r.v. from $\left|h_{n,1}\right|,\,...,\,\left|h_{n,N}\right|$
that are smaller than $\varepsilon$ has a binomial distribution with
$p_{n}=\Pr\left(\left|h_{n,1}\right|<\varepsilon\right)$. We use
the Chernoff-Hoeffding Theorem \cite{Hoeffding1963} as a tail bound
for $\frac{M_{N}}{N}<p_{n}$. By the assumption on $M_{N}$, $\frac{M_{N}}{N}<p_{n}$
holds for all $N>N_{1}$ for some large enough $N_{1}$, and so
\[
\Pr\left(N_{\varepsilon,n}\leq M_{N}\right)\leq\exp\left(-ND\left(\frac{M_{N}}{N}\Vert p_{n}\right)\right)
\]
where $D(q\Vert p)=q\ln\frac{q}{p}+\left(1-q\right)\ln\frac{1-q}{1-p}$
, and in our case

\begin{multline*}
D\left(\frac{M_{N}}{N}\Vert p_{n}\right)=\frac{M_{N}}{N}\ln\frac{M_{N}}{Np_{n}}+\left(1-\frac{M_{N}}{N}\right)\ln\frac{1-\frac{M_{N}}{N}}{1-p_{n}}
\end{multline*}
for which
\begin{multline*}
\underset{N\rightarrow\infty}{\lim}D\left(\frac{M_{N}}{N}\Vert p_{n}\right)=-\underset{N\rightarrow\infty}{\lim}\frac{\ln\frac{N}{M_{N}}}{\frac{N}{M_{N}}}\\
-\ln p_{n}\underset{N\rightarrow\infty}{\lim}\frac{M_{N}}{N}+\ln\left(\frac{1}{1-p_{n}}\right)\underset{N\rightarrow\infty}{\lim}\left(1-\frac{M_{N}}{N}\right)\\
+\underset{N\rightarrow\infty}{\lim}\left(1-\frac{M_{N}}{N}\right)\ln\left(1-\frac{M_{N}}{N}\right)=\ln\left(\frac{1}{1-p_{n}}\right)
\end{multline*}
So for large enough $N$ the inequality $D\left(\frac{M_{N}}{N}\Vert p_{n}\right)\geq\ln\left(\frac{1}{1-p_{n}}\right)-\ln\left(\frac{1}{1-p_{n}^{2}}\right)$
holds and hence we get the following upper bound
\begin{multline*}
\Pr\left(N_{\varepsilon,n}\leq M_{N}\right)\leq\exp\left(-ND\left(\frac{M_{N}}{N}\Vert p_{n}\right)\right)\\
\leq\left(1-p_{n}\right)^{N}\left(\frac{1}{1-p_{n}^{2}}\right)^{N}=\left(\frac{1}{1+p_{n}}\right)^{N}\rightarrow0
\end{multline*}
Clearly, if there are at least $M_{N}$ successes then the $M_{N}$
smallest variables among $\left|h_{n,1}\right|,\,...,\,\left|h_{n,N}\right|$
are smaller than $\varepsilon$. Consequently, using the union bound
we get
\begin{multline*}
\underset{N\rightarrow\infty}{\lim}\Pr\left(\underset{n}{\max\,}\left|h_{n,(m)}\right|>\varepsilon\right)=\\
\underset{N\rightarrow\infty}{\lim}\Pr\left(\bigcup_{n=1}^{N}\left\{ \left|h_{n,(m)}\right|>\varepsilon\right\} \right)\leq\underset{N\rightarrow\infty}{\lim}\sum_{n=1}^{N}\frac{1}{\left(1+p_{n}\right)^{N}}=0
\end{multline*}
for each $m\leq M_{N}$ and each $\varepsilon>0$, and we reached
our conclusion.
\end{IEEEproof}
Since $\frac{1}{N}\sum_{n}w_{n}u_{n}\leq w_{max}\underset{n}{\max\,}u_{n}$,
it follows from the lemma above that the users that are assigned one
of their $M_{N}$ worst channel coefficients have an average utility
that converges to zero in probability. To evaluate the performance
of the worst PNE of the Naive-FSIG, we need to know how many users
can be assigned such a bad channel. Unfortunately, there is an $M_{N}$
s.t. $\underset{N\rightarrow\infty}{\lim}\frac{M_{N}}{N}=0$, for
which there exists a permutation between users and channels such that
each user gets one of his $M_{N}$ worst channel coefficients. Even
worse, there are actually many such permutations. Our result is based
on the following theorem.
\begin{thm}[Frieze \& Melsted \cite{Frieze2012}]
\label{thm:-Let-} Let $\varGamma$ be a bipartite graph chosen uniformly
from the set of graphs with bipartition $L,R$, $|L|=n,|R|=m$ s.t.
each vertex of $L$ has degree $d\geq3$ and each vertex of $R$ has
degree at least two. Then with high probability the maximum matching
in $\varGamma$ is with size $min\left\{ m,n\right\} .$
\end{thm}

\begin{lem}
\label{Existence of a Perfect Matching}Assume that $\left\{ h_{n,m,k}\right\} $
are independent and $h_{n,1},\,...,\, h_{n,N}$ are identically distributed
for each $n$. Let $\mathcal{M}_{n}=\left\{ k\,|\,\left|h_{n,k}\right|\leq\left|h_{n,(M_{N})}\right|\right\} $.
If $M_{N}\geq(1+\varepsilon)\ln(N)$ for some $\varepsilon>0$, then
the probability that a perfect matching exists between users and channels
s.t. each $n\in\mathcal{N}$ gets a channel from $\mathcal{M}_{n}$
approaches 1 as $N\rightarrow\infty$.\end{lem}
\begin{IEEEproof}
The degree of each user node is exactly $M_{N}$. We want to bound
the probability of the event that there exists a channel with degree
zero or one. Due to the i.i.d channel coefficients of each user, the
probability that user $n$ is not connected to channel $k$ is given
by 
\[
\Pr\left(k\notin\mathcal{M}_{n}\right)=\frac{\left(\begin{array}{c}
N-1\\
M_{N}
\end{array}\right)}{\left(\begin{array}{c}
N\\
M_{N}
\end{array}\right)}=\frac{\frac{(N-1)!}{(N-1-M_{N})!}}{\frac{N!}{(N-M_{N})!}}=1-\frac{M_{N}}{N}
\]
Since channel coefficients of different users are independent, the
probability that the degree of vertex $k$ is less than two is given
by the binomial distribution of the number of users who prefer channel
$k$, with $p=\frac{M_{N}}{N}$

\[
\Pr\left(\deg(k)<2\right)=\left(1-\frac{M_{N}}{N}\right)^{N}+N\frac{M_{N}}{N}\left(1-\frac{M_{N}}{N}\right)^{N-1}
\]
So by the union bound on the channel vertices and for $M_{N}\geq(1+\varepsilon)\ln(N)$
with some $\varepsilon>0$ we get, for $\delta>0$ and large enough
$N$, that

\begin{multline*}
\Pr\left(\underset{k}{\min}\deg(k)<2\right)\leq\\
N\left(1-\frac{M_{N}}{N}\right)^{N}+NM_{N}\left(1-\frac{M_{N}}{N}\right)^{N-1}\leq\\
(1+\delta)\left(Ne^{-M_{N}}+NM_{N}e^{-M_{N}}\right)\\
\leq(1+\delta)\frac{(1+\varepsilon)\ln(N)+1}{N^{\varepsilon}}
\end{multline*}
which goes to zero as $N\rightarrow\infty.$ We know from Theorem
\ref{thm:-Let-} that given $\underset{k}{\min}\left(\deg(k)\right)\geq2$,
the probability that a perfect matching exists approaches 1 as $N\rightarrow\infty$
so by combining these results we obtain our conclusion. 
\end{IEEEproof}
The condition $M_{N}\geq(1+\varepsilon)\ln(N)$ was necessary to ensure
that with high probability, no channel node degree is smaller than
two. This large user nodes degree has its own major effect on the
equilibria as well.
\begin{thm}[{Marshall Hall Jr \cite[Theorem 2]{HallJr1948}}]
\label{Hall Jr} Suppose that $A_{1},A_{2},...,A_{N}$ are the finite
sets of desirable resources, i.e. user $n$ desires resource $a$
if and only if $a\in A_{n}$. If there exists a perfect matching between
users and resources and $\left|A_{n}\right|\geq M$ for $n=1\,,...,\, N$
where $M<N$, then the number of perfect matchings is at least $M!$.
\end{thm}
Joining together Theorem \ref{Hall Jr}, Lemma \ref{NE are Permutations},
Lemma \ref{Worst Channels are Asymptotically Worthless} and Lemma
\ref{Existence of a Perfect Matching} we arrive at the following
theorem, by choosing $M_{N}=N^{\mu}$ for some $\mu<1$.
\begin{thm}
Assume that $\left\{ h_{n,m,k}\right\} $ are independent and $\left|h_{n,1}\right|,\,...,\,\left|h_{n,N}\right|$
are identically distributed for each $n$, with continuous distribution
$F_{n}\left(x\right)$, s.t. $F_{n}\left(x\right)>0$ for all $x>0$.
Also assume that there exist $w_{min},w_{max}>0$, s.t. $w_{min}\leq w_{n}\leq w_{max}$
for all $n$. If $\frac{1}{N_{0}}\underset{k}{\min}\left|h_{n,k}\right|^{2}>\underset{l}{\max}\frac{\left|h_{n,l}\right|^{2}}{N_{0}+\underset{m}{\min}\left(\left|h_{m,n,l}\right|^{2}P_{m}\right)}$
holds for each $n$, then for all $\mu<1$, there are at least $\left(N^{\mu}\right)!$
PNE s.t. $\frac{1}{N}\sum_{n}w_{n}u_{n}\rightarrow0$ in probability
as $N\rightarrow\infty$. Specifically, The PPoA of the Naive-FSIG
approaches infinity in probability as $N\rightarrow\infty$.
\end{thm}

\section{The M Frequency-Selective Interference Game }

In this section, we want to exploit the degrees of freedom in choosing
the utility function. Inspired by the hazards demonstrated by the
naive game, we propose a new game formulation. The greediness of the
users is moderated by an a-priori agreement to limit the utility of
each user to be greater than zero only for his $M$ best channels,
and equal for them.
\begin{defn}
\textbf{The M Frequency-Selective Interference Game (M-FSIG)} is a
normal-form game with parameter $M>0$, $N$ users as players, where
each has $A_{n}=\left\{ 1,2,...,K\right\} $ as a strategy space.
The utility function for player $n$ is

\[
u_{n}\left(\mathbf{a}\right)=\left\{ \begin{array}{cc}
\log_{2}\left(1+\frac{P_{n}|h_{n,(N-M+1)}|^{2}}{N_{0}+I_{n,a_{n}}\left(\mathbf{a}_{-n}\right)}\right) & \frac{\left|h_{n,a_{n}}\right|}{\left|h_{n,(N-M+1)}\right|}\geq1\\
0 & else
\end{array}\right.
\]

\end{defn}
Define the set of indexes of the $M$ best channel coefficients of
user $n$ by $\mathcal{M}_{n}=\left\{ k\,|\,\frac{\left|h_{n,k}\right|}{\left|h_{n,(N-M+1)}\right|}\geq1\right\} .$
Note that because $\left|h_{n,(N-M+1)}\right|>0$ for each $n\in\mathcal{N}$
with probability 1, user $n$ will never choose a channel outside
$\mathcal{M}_{n}$. Also note that due to the replacement of $h_{n,a_{n}}$
by $h_{n,(N-M+1)}$ in the utility, maximizing $u_{n}$ is equivalent
to minimizing $I_{n,a_{n}}.$ Hence in the M-FSIG each user $n\in\mathcal{N}$
accesses only channels in $\mathcal{M}_{n}$ and prefers those with
smaller interference. 

The identification of $\mathcal{M}_{n}$ is superior both in performance
and practice over evaluating all the channels that are better than
some threshold as was done in \cite{Naparstek2013,Naparstek2014}.
If this threshold is constant with respect to $N$ a significant rate
loss may occur due to truncation, because the expected value of the
best channel coefficients grows with $N$. If the threshold is dependent
on $N$, this dependence is dictated by the fading distribution, which
is not known to the users. 

We will show that the M-FSIG has asymptotically only good PNE in any
interference regime. For this reason, in this game, the convergence
to some PNE is sufficient to provide good global performance. Furthermore,
it only requires each user to track a small number of channels (e.g.
$O\left(\ln N\right)$ instead of $N$).

\subsection{Existence of an Asymptotically Optimal Permutation Equilibrium }

In this subsection we establish that as $N\rightarrow\infty$ the
probability that an asymptotically optimal PNE exists approaches 1. 

The existence result in Lemma \ref{Existence of a Perfect Matching}
(and Theorem \ref{Hall Jr}) of permutations where each user $n$
gets a channel from $\mathcal{M}_{n}$ is of course unaffected by
how we choose the $M$ members of the set $\mathcal{M}_{n}$. With
$\mathcal{M}_{n}$ as defined for the M-FSIG we get the following
corollary.
\begin{cor}
Assume that $\left\{ h_{n,m,k}\right\} $ are independent and $h_{n,1},\,...,\, h_{n,N}$
are identically distributed for each $n$. If the M-FSIG parameter
is chosen s.t. $M\geq(1+\varepsilon)\ln(N)$ for some $\varepsilon>0$,
then the probability that the M-FSIG has at least $M!$ PNE that are
permutations of users to channels, s.t. user $n$ gets a channel from
$\mathcal{M}_{n}$ for each $n\in\mathcal{N}$, approaches 1 as $N\rightarrow\infty$.
\end{cor}
Finally, we state the result that shows that this set of permutation
equilibria are indeed asymptotically optimal. This result depends
on the statistics of the fading coefficients, and here we choose to
present the common case of Rayleigh fading, although the same is true
for a broad class of fading distributions. Details will be given in
\cite{Bistritz}. 
\begin{thm}
\label{Max-Min Fairness Equilibria}Assume that $\left\{ h_{n,m,k}\right\} $
are independent and $|h_{n,1}|,...,|h_{n,N}|$ are i.i.d Rayleigh
distributed variables for each $n$. If $\underset{N\rightarrow\infty}{\lim}\frac{M}{N^{\mu}}=0$
for $\mu<1$ and $\underset{N\rightarrow\infty}{\lim}M=\infty$, then
there exist at least $M!$ PNE for the M-FSIG for which $\underset{n\in\mathcal{N}}{\min}\frac{\log_{2}\left(1+\frac{P_{n}}{N_{0}}|h_{n,a_{n}^{*}}|^{2}\right)}{\log_{2}\left(1+\frac{P_{n}}{N_{0}}|h_{n,(N)}|^{2}\right)}\rightarrow1$
in probability as $N\rightarrow\infty$. 
\end{thm}

\subsection{Non-Existence of Bad Equilibria }

It turns out that the asymptotically optimal permutation PNE are typical
equilibria for this game; in other words all other equilibria are
almost a permutation and hence have the same asymptotic performance.
This property eases the requirements for the dynamics and allows simpler
convergence with good performance. We define a sharing user as a user
who is in the same channel with at least one more user. 
\begin{thm}
Assume that $\left\{ h_{n,m,k}\right\} $ are independent and $h_{n,1},\,...,\, h_{n,N}$
are identically distributed for each $n$. Suppose that $M\geq(1+\varepsilon)\ln(N)$
for some $\varepsilon>0$. If \textbf{$\mathbf{a^{*}}$} is a PNE
of the M-FSIG with $N_{c}$ sharing users, then $\frac{N_{c}}{N}\rightarrow0$
in probability as $N\rightarrow\infty.$ \end{thm}
\begin{IEEEproof}
The proof is based on the fact that the larger the number of sharing
users, the larger the number of empty channels and hence the probability
that none of these empty channels is in $\mathcal{M}_{n}$ for some
$n$ decreases with $N$. Details are omitted due to page constraints
and will be given in \cite{Bistritz}.
\end{IEEEproof}
The weighted sum-rate of the non-sharing users approaches the optimal
one, and almost all users are non-sharing users. Nevertheless, the
sharing users do not necessarily suffer from poor conditions - they
are in their minimal interference channel out of an increasing (with
$N$) amount of good channels. 

This result, aided by the existence of an asymptotically good permutation
PNE from the last section, leads to the following satisfactory property
of the M-FSIG that holds for any interference regime. 
\begin{cor}
Assume that $\left\{ h_{n,m,k}\right\} $ are independent and $|h_{n,1}|,...,|h_{n,N}|$
are i.i.d Rayleigh r.v. for each $n$, and that $w_{min}\leq w_{n}\leq w_{max}$
with some $w_{min},w_{max}>0$, for each $n$. If $M=(1+\varepsilon)\ln\left(N\right)$
for some $\varepsilon>0$, then each PNE $\mathbf{a}^{*}$satisfies
\[
\frac{\sum_{n=1}^{N}w_{n}\log_{2}\left(1+\frac{P_{n}\left|h_{n,a_{n}^{*}}\right|^{2}}{N_{0}+I_{n,a_{n}^{*}}}\right)}{\sum_{n=1}^{N}w_{n}\log_{2}\left(1+\frac{P_{n}}{N_{0}}\left|h{}_{n,(N)}\right|^{2}\right)}\rightarrow1
\]
in probability as $N\rightarrow\infty$. Consequently, the PPoA of
the M-FSIG converges to 1 in probability as $N\rightarrow\infty$.
\end{cor}
Note that because for $M=(1+\varepsilon)\ln\left(N\right)$ some of
the PNE of the M-FSIG are permutations, the above corollary suggests
that the ratio of the weighted sum-rate of the optimal permutation
to $\sum_{n=1}^{N}w_{n}\log_{2}\left(1+\frac{P_{n}}{N_{0}}\left|h{}_{n,(N)}\right|^{2}\right)$
converges to 1 in probability as $N\rightarrow\infty$.

\section{Modified Fictitious Play}

In order to apply game theory algorithmically, an equilibrium analysis,
although necessary, is not enough. A learning algorithm that each
user can implement that leads to convergence to equilibrium is a crucial
element. It should be emphasized that the performance of such an algorithm
is already known from the equilibrium analysis. Therefore the algorithm
that we are looking for is not tailored to our specific problem but
rather has general properties of convergence to NE. One of the best
known candidates for this task is the Fictitious Play (FP) algorithm
\cite{Brown1951}. In FP, each player keeps the empirical mean vectors
of the frequencies each other player has played his actions, and plays
his best response to this fictitious strategy. Alternatively a player
can keep one empirical mean vector of the frequencies each strategy
profile of its rivals has been played (joint strategy fictitious play).
The simple relation of FP convergence to NE is summarized in the following
proposition.
\begin{prop}[\cite{Levin}]
 Let $N$ players play according to the FP algorithm. 
\begin{enumerate}
\item If a PNE is attained at $t_{0}$ it will be played for all $t>t_{0}$
and the frequency vectors will converge to it.
\item If FP frequency vectors converge, they must converge to some NE (maybe
mixed). 
\item If $\mathbf{a\in}A_{1}\mathbf{\times...\times}A_{N}$ is played for
every $t>t_{1}$ then $\mathbf{a}$ is a PNE. 
\end{enumerate}
\end{prop}
Although its strong connection to NE, FP is not guaranteed to converge
at all. Convergence has been proven only for some special games that
do not include our game. Even if FP converges, it may be to a mixed
NE and this is undesirable as was mentioned above. Furthermore, a
common problem with implementing FP is the information it requires.
In a wireless network, not only does a user have hardly any information
about the previous action of each other user, but he also barely knows
how many users there are. Fortunately, in our game the effect of the
other users on the utility can be measured by measuring the interference. 

To adjust the FP to the wireless environment we propose to modify
it such that each user keeps track of a fictitious utility vector
instead of the empirical mean vector of the rivals strategy profiles.
We denote the fictitious utility for user $n$ in channel $k$ at
time $t$ by $\overline{U}_{n,k}(t)$. The fictitious utility is updated
according to the following rule 
\[
\overline{U}_{n,k}(t)=(1-\alpha)\overline{U}_{n,k}(t-1)+\alpha u_{n,k}(t)
\]
with some step size $0<\alpha\leq1$. To prevent mixed NE we suggest
a constant step size instead of the common $\alpha=\frac{1}{t+1}$
that makes $\overline{U}_{n,k}$ the empirical mean utility. Note
that $\alpha=1$ corresponds to the best-response dynamics.

Additionally, we provide a simple mechanism to improve the chances
of convergence to a PNE. The strategy profile determines the interference,
but knowing the interference will not reveal the strategy profile.
Nevertheless, the continuity of the random channel gains suggests
that for each user, the interference vector is different for different
strategy profiles with probability 1. Hence users can detect that
two strategy profiles are different based on their measured interference.
If a PNE is reached it is played repeatedly, so we can exploit this
fact and let the users check for convergence to a PNE after enough
time, and set their fictitious utility to zero if a PNE has not been
reached. 

The Modified FP is described in detail in the Algorithm 1 table, and
its properties are summarized in the next proposition.
\begin{prop}
Let $N$ players play according to the Modified FP algorithm. 
\begin{enumerate}
\item If $\alpha=\frac{1}{t+1}$, then the dynamics of the joint strategy
FP where each player has perfect information are identical to those
of the Modified FP.
\item Assume a constant $\alpha$. If a PNE is attained at $t_{0}$ then
it will be played for all $t>t_{0}$ and if the fictitious utility
vectors converge, then the resulting strategy profile is a PNE.
\end{enumerate}
\end{prop}
\begin{IEEEproof}
Assume we are at turn $t=T$ and define $p_{i}=\sum_{t=1}^{T}\frac{I(\mathbf{a}_{-n}(t)=\mathbf{a}_{i,-n})}{T}$
for the rivals strategy profile $\mathbf{a}_{i,-n}$, where $I$ is
the indicator function. For $\alpha=\frac{1}{t+1}$ the equivalence
of the algorithms follows immediately from the identity 
\begin{multline*}
\sum_{i}p_{i}u_{n}(a_{n},\mathbf{a}_{i,-n})=\\
\sum_{i}p_{i}\log_{2}\left(1+\frac{P_{n}|h_{n,(N-M+1)}|^{2}}{N_{0}+I_{n,a_{n}}(\mathbf{a}_{i,-n})}\right)\\
=\frac{1}{T}\sum_{t=1}^{T}\log_{2}\left(1+\frac{P_{n}|h_{n,(N-M+1)}|^{2}}{N_{0}+I_{n,a_{n}}\left(\mathbf{a}_{-n}\left(t\right)\right)}\right)
\end{multline*}
because $\sum_{i}p_{i}u_{n}(a_{n},\mathbf{a}_{i,-n})$ is the mean
empirical utility for $a_{n}$ according to the fictitious rivals
profile. Consider next the case of a constant $\alpha$. If a PNE
$\mathbf{a^{*}}$ is attained at $t_{0}$ then $a_{n}^{*}(t_{0})=\arg\underset{k}{\max}\, u_{n,k}(t_{0})$
and $a_{n}^{*}(t_{0})=\arg\underset{k}{\max}\,\overline{U}_{n,k}(t_{0}-1)$
for each $n\in\mathcal{N}$. Considering the update rule and because
$a_{n}^{*}(t_{0})=\arg\underset{k}{\max\,}\overline{U}_{n,k}(t_{0}-1)=\arg\underset{k}{\max\,}u_{n,k}(t_{0})$%
\footnote{For the proof it is enough to break ties in $\overline{U}_{n,k}(t)$
by choosing the previous action if it is maximal, otherwise break
ties arbitrarily. For $t=0$ step (d) of the Modified FP suggests
breaking ties at random.%
} we get 
\begin{multline*}
\arg\underset{k}{\max}\,(1-\alpha)\overline{U}_{n,k}(t_{0}-1)+\arg\underset{k}{\max}\,\alpha u_{n,k}(t_{0})=\\
\arg\underset{k}{\max\,}\overline{U}_{n,k}(t)=a_{n}^{*}(t_{0}+1)
\end{multline*}
and so on, for each $t>t_{0}.$ If the fictitious utility vectors
converge, then $\underset{t\rightarrow\infty}{\lim}\overline{U}_{n,k}(t)$
exists and is finite for each $k$ and $n$. From the update rule
we get $\alpha\underset{t\rightarrow\infty}{\lim}\overline{U}_{n,k}(t)=\alpha\underset{t\rightarrow\infty}{\lim}u_{n,k}(t)$
for each $n,k$ which means $\underset{t\rightarrow\infty}{\lim}\overline{U}_{n,k}(t)=\underset{t\rightarrow\infty}{\lim}u_{n,k}(t)$
for constant $\alpha$. Consequently, for all $t>t_{1}$ for some
large enough $t_{1}$, $a_{n}(t)=\arg\underset{k}{\max}\,\overline{U}_{n,k}(t)=\arg\underset{k}{\max}\, u_{n,k}(t)$
for each $n\in\mathcal{N}$ and hence $\mathbf{a}$ is a PNE. 
\end{IEEEproof}
In the next section we show that in our case there is indeed convergence
to PNE, and a very fast one. 

\begin{algorithm}[H]
\protect\caption{Modified Fictitious Play}

\begin{enumerate}
\item \textbf{Initialization} - Choose some $0<\alpha\leq1$ and $\tau>0$.
Each user initializes his fictitious utility - $\bar{U}_{n,k}(0)=0$
for each $k\in\mathcal{M}_{n}$, where $\mathcal{M}_{n}$ is the set
of his $M$ best channels (interference free).\\

\item \textbf{For t=1,...T and for each user n=1,...N do}

\begin{enumerate}
\item Choose a transmission channel 
\[
a_{n}(t)=\arg\underset{k}{\max}\,\overline{U}_{n,k}(t-1)
\]

\item Sense the interference. For each $k\in\mathcal{M}_{n}$
\[
I_{n,k}(t)=\underset{m|a_{m}(t)=k}{\sum}|h_{m,n,k}|^{2}P_{m}
\]

\item Update fictitious utilities. For each $k\in\mathcal{M}_{n}$ 
\[
\overline{U}_{n,k}(t)=(1-\alpha)\overline{U}_{n,k}(t-1)+\alpha u_{n,k}(t)
\]
where 
\[
u_{n,k}(t)=\log_{2}\left(1+\frac{P_{n}|h_{n,(N-M+1)}|^{2}}{N_{0}+I_{n,k}(t)}\right)
\]

\item (optional) Check for convergence to a PNE. If $t=\tau$ and $I_{n,k}(t)\neq I_{n,k}(t-1)$
for some $k\in A_{n}$ then return to step 1, i.e. $t=0$.\end{enumerate}
\end{enumerate}
\end{algorithm}

\section{Simulation Results}

Our analysis is probabilistic and asymptotic with the number of users.
Thus, we carried out some simulations to ensure that the asymptotic
effects take place for reasonable $N$ values. In fact, the situation
for some finite $N$ tends to be much more optimistic than the lower
bounds we used in most of our proofs. 

In our simulations we used a Rayleigh fading network; i.e. $\left\{ \left|h_{m,n,k}\right|\right\} $
are i.i.d Rayleigh random variables. Hence $\left\{ \left|h_{m,n,k}\right|^{2}\right\} $
are i.i.d exponential random variables with parameter $\lambda$,
which is chosen to be $\lambda=1$ so all the exponential variables
have unit variance. Unless specified otherwise, the transmission powers
were chosen such that the mean SNR for each link, in the absence of
interference, is 20{[}dB{]}. Users play according to the Modified
FP algorithm from last section including step (d) with $\tau=60$
and $\alpha=0.5$. 

In Fig. 1 we present the convergence of the Modified FP in two different
network realizations, for $N=K=100$. We can see that convergence
is very fast and occurs within 100 iterations. The upper figure is
for $M=9$, where the ratio of the sum of achievable rates to that
of an optimal allocation is close to 1, and the ratio of the minimal
achievable rate is a bit smaller. This corresponds to a convergence
to one of the permutation PNE; hence, there are no sharing users.
The lower figure shows another realization for both $M=7,14$. The
sum-rate ratio to optimal is still close to 1, as predicted by the
converging PPoA, but the minimal rate is significantly lower. The
minimal rate is experienced by one of the four sharing users in this
case. Nevertheless, choosing $M=14$ results in a negligible reduction
in the sum-rate but significantly improves the minimum rate due to
the convergence to a permutation with no sharing users. We can also
see the benefits of step (d) of the Modified FP, which leads to a
detection of non-convergence by the users at $t=60$. This indeed
results in a convergence to a PNE after the users selected initial
channels at random again. 

In Fig. 2 we show the effect of the number of users on the rates,
with $M=\left\lceil 3\ln\left(N\right)\right\rceil $. We present
the average and minimal achievable rates, compared to the sum-rate
optimal permutation allocation and random permutation allocation average
and minimal achievable rates. The benefit over a random permutation
is significant, especially in terms of the minimal rate. The rates
increase is due to the growing expected value of the best channel
coefficients for each user. This phenomenon (multi-user diversity)
of course does not take place for a random permutation. In a random
permutation the average user gets his median channel coefficient,
and the minimal allocated channel coefficient has a decreasing expectation.
The standard deviations of the mean rates are small as expected from
the similarity of all NE, and the standard deviations of the minimal
rate are higher due to changing number of sharing users between different
realizations. 

In Fig. 3 we show the effect of the mean SNR on the gain of our algorithm
over a random permutation allocation, compared to the optimal permutation
allocation. This simulation is for $N=200$ and $M=\left\lceil 3\ln\left(N\right)\right\rceil $.
As the mean SNR grows larger, the logarithmic behavior of the achievable
rate causes the rate difference between two given channel coefficients
to be smaller. At the same time, the rate difference between an occupied
channel and a free one grows and hence the number of sharing users
drops (7.68 at $SNR=-10[dB]$ and 1.2 at $SNR=25[dB]$).

\begin{figure}[tbh]
\begin{minipage}[t]{1\columnwidth}%
\includegraphics[clip,width=9cm,height=4.5cm]{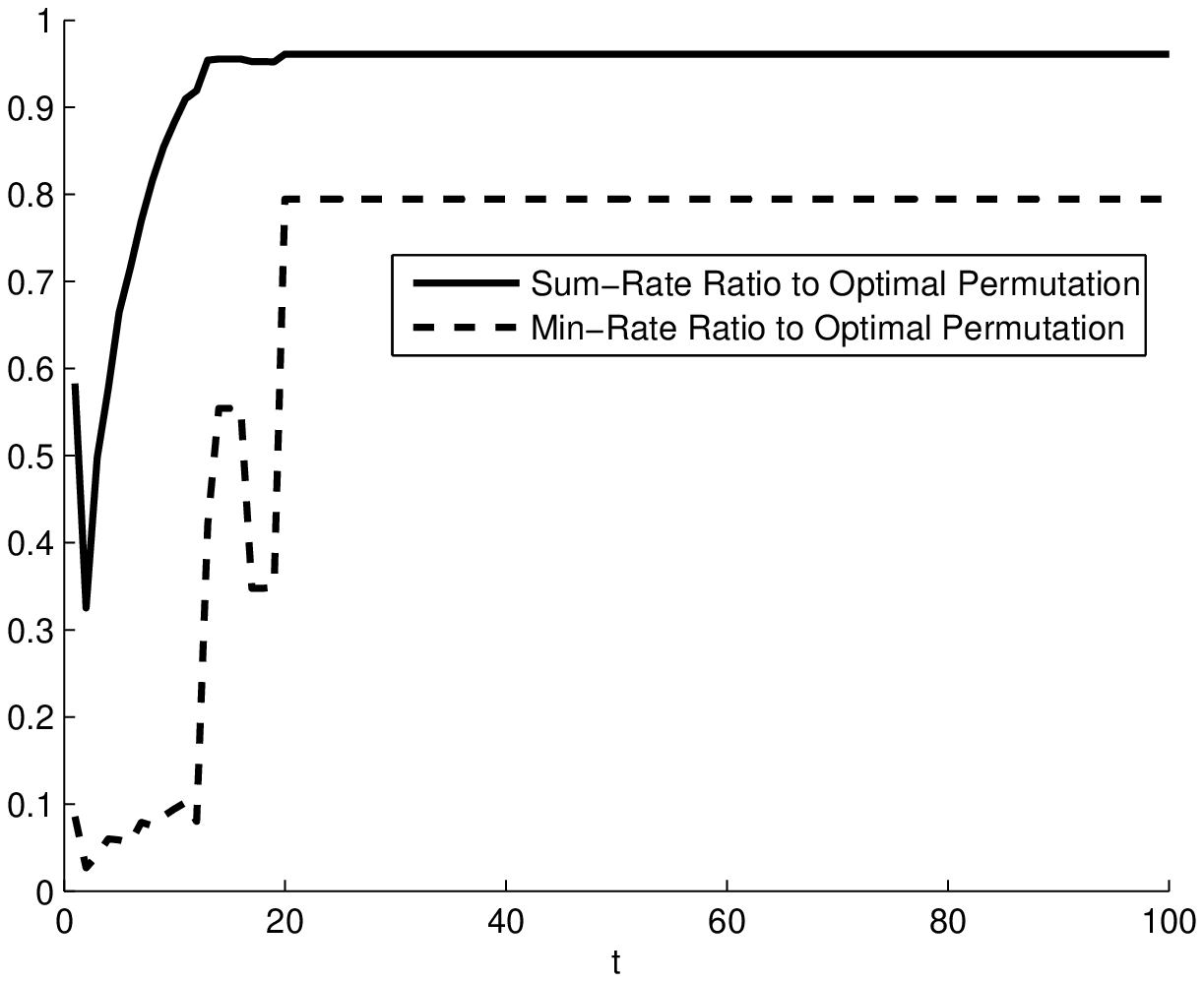}%
\end{minipage}

\begin{minipage}[t]{1\columnwidth}%
\includegraphics[clip,width=9cm,height=4.5cm]{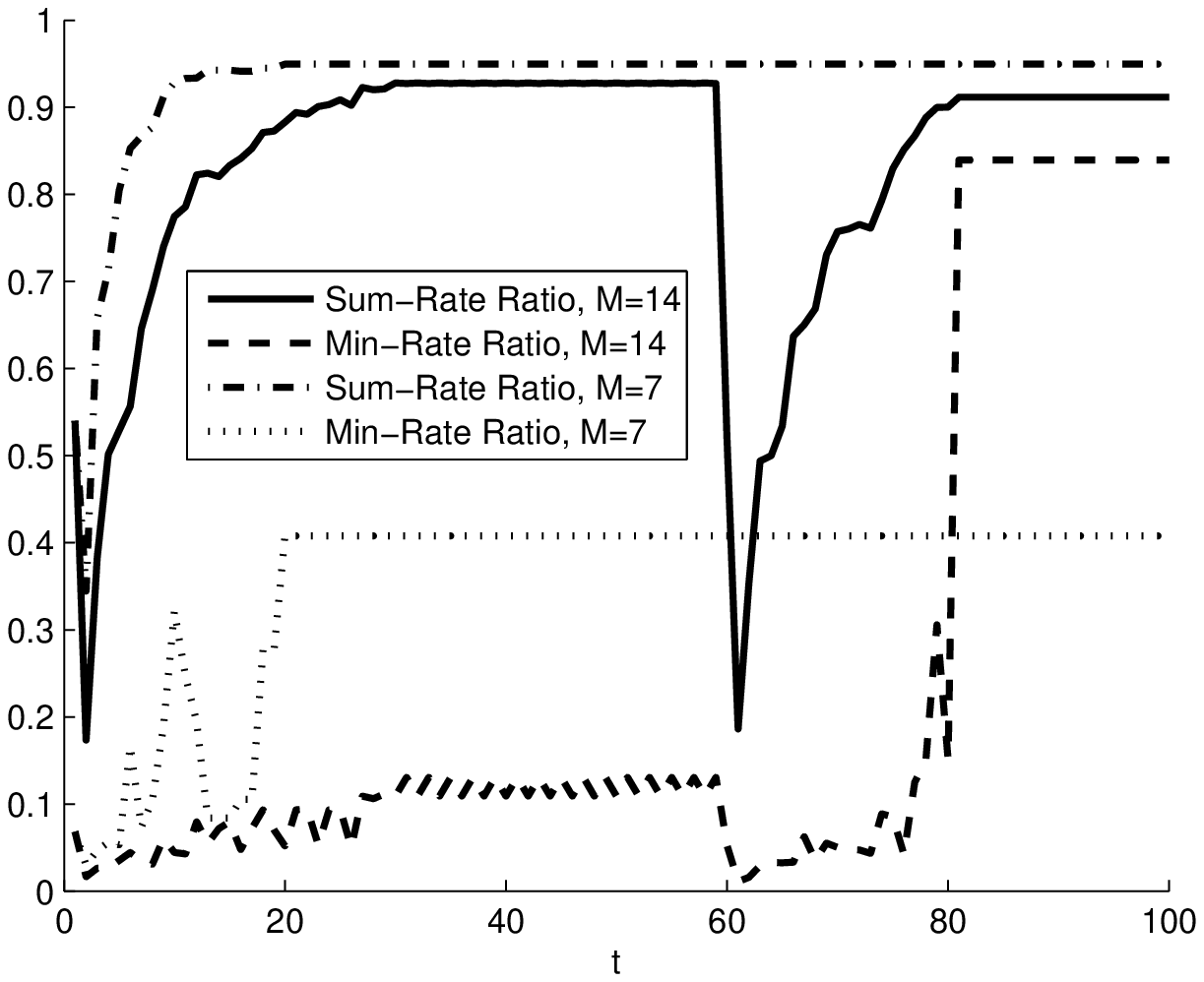}%
\end{minipage}

\protect\caption{Sum-rate and min-rate compared to the optimal permutation allocation
sum-rate for two different realizations}
\end{figure}

\begin{figure}[tbh]
\begin{minipage}[t]{1\columnwidth}%
\includegraphics[clip,width=9cm,height=4.5cm]{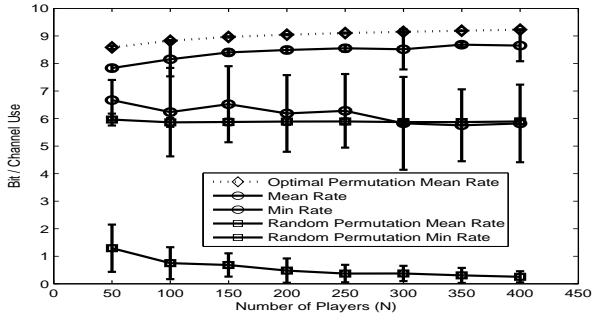}%
\end{minipage}

\protect\caption{Rates as a function of $N$ averaged over 50 realizations}
\end{figure}

\begin{figure}[tbh]
\begin{minipage}[t]{1\columnwidth}%
\includegraphics[clip,width=9cm,height=4.5cm]{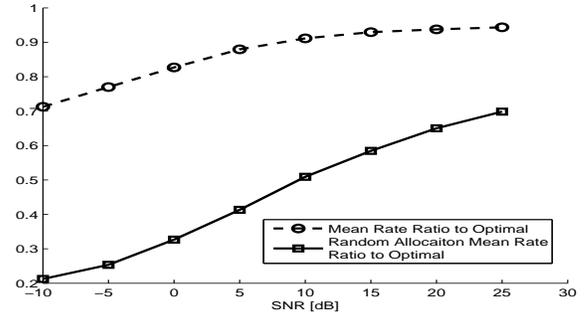}%
\end{minipage}

\protect\caption{Mean rates as a function of the mean SNR, averaged over 50 realizations}
\end{figure}

\section{Conclusion}

In this paper we analyzed, using asymptotic probabilistic tools, two
game formulations for the distributed channel allocation problem in
the frequency-selective interference channel. The performance metric
was the weighted sum of achievable rates when treating interference
as noise.

First we presented a naive non-cooperative game (Naive-FSIG) and showed
that with strong enough interference it has $\varOmega\left(\left(N^{\mu}\right)!\right)$,
for all $\mu<1$, bad pure NE, where $N$ is the number of users.

We then proposed an enhanced non-cooperative game formulation (M-FSIG)
based on an a-priori agreement between users to limit their utility
to be greater than zero only for their $M$ best channels, with the
same value for those channels. We proved that for many fading distributions
(including Rayleigh fading), our game has a pure price of anarchy
that approaches 1 as $N\rightarrow\infty$ in any interference regime.
For some fixed $N$ the introduced parameter $M$ can be chosen to
compromise between sum-rate and fairness. This game enables a fully
distributed implementation that achieves close to optimal performance
without resorting to coordinated solutions.

Due to the almost completely orthogonal transmissions in equilibria
our allocation algorithm is more suitable for the medium-strong interference
regime.

We also proposed a modified Fictitious Play algorithm and showed through
simulations that it converges very fast to the proven pure NE. The
fast convergence enables frequent runs of the algorithm in the network,
which results in maintaining the multi-user diversity in a dynamic
fading environment. 

The simplicity and high performance of our algorithm make it an appealing
base for a dynamic channel access protocol for distributed networks.

\section*{Acknowledgment}

This research was supported by the Israeli Science Foundation, under
grant 903/2013.

\appendices{}

\bibliographystyle{IEEEtran}
\bibliography{\string"Asymptotically Optimal Distributed Channel Allocation - a Game-Theoretic Approach\string"}

\end{document}